\newcommand{\Mpc}{\mathrm{Mpc}}

\newcommand{\dd}{\mathrm{d}}

\newcommand{\fNL}{f_{\mathrm{NL}}}
\newcommand{\fNLeq}{f_{\mathrm{NL}}^{\mathrm{eq}}}

\documentclass[cits]{PoS}
\usepackage{amsmath}

\title{The Effects of Primordial Non-Gaussianity on Giant-Arc Statistics:  A Scale Dependent Example}

\ShortTitle{Giant-Arc Statistics with Primordial non-Gaussianity}

\author{\speaker{Anson D'Aloisio}\\
        Department of Astronomy \& Texas Cosmology Center, University of Texas at Austin \\
        Department of Physics, Yale University
\\        E-mail: \email{anson@astro.as.utexas.edu}}

\author{Priyamvada Natarajan\\
         Department of Physics, Yale University \\
         	Department of Astronomy, Yale University\\
        E-mail: \email{priyamvada.natarajan@yale.edu}}

\abstract{In a recently published article, we quantified the impact of primordial non-Gaussianity on the probability of giant-arc formation.  In that work, we focused on the local form of non-Gaussianity and found that it can have only a modest effect given the most recent constraints from Cosmic Microwave Background (CMB) measurements.  Here, we present new calculations using a parameterization of scale-dependent non-Gaussianity in which the primordial bispectrum has the equilateral shape and the effective $\fNL$ parameter depends on scale.  We find that non-Gaussianity of this type can yield a larger effect on the giant-arc abundance compared to the local form due to both the scale dependence and the relatively weaker constraints on the equilateral shape from CMB measurements.  In contrast to the maximum $\sim40\%$ effect (within the latest CMB constraints) previously found for the local form, we find that the predicted giant-arc abundance for the scale-dependent equilateral form can differ by a factor of a few with respect to the Gaussian case.    }

\FullConference{Frank N. Bash Symposium New Horizons In Astronomy,\\
		October 9-11, 2011\\
		Austin Texas}

\begin{document}

The formation of giant arcs by strong gravitational lensing is reserved for the most massive collapsed structures whose statistical properties are sensitive to the expansion history and initial conditions of the Universe.  Since the frequency of giant-arc formation depends on the abundance and characteristics of galaxy-clusters roughly half-way to the sources, it has long been recognized as a potentially rich source of information.  

At the same time, the interplay between cosmological effects, cluster physics, and the source population makes their disentanglement non-trivial. The difficulties have been brought to light for over a decade following the initial claim of  \cite{1998A&A...330....1B} that $\Lambda$CDM predicted approximately an order of magnitude fewer arcs than seen in observations.   This claim stimulated a large amount of work towards understanding the most important characteristics of arc-producing clusters,  how they may differ from the general cluster population, and the role of source characteristics in giant-arc production (see references in the introduction of \cite{2011MNRAS.415.1913D}).

Despite such extensive efforts, the status of the giant-arc problem still remains unclear (see references in \cite{2011MNRAS.415.1913D}). It is still possible that the cosmological model may have at least a partial role to play.  Motivated by this fact, and the recent interest that structure formation with primordial non-Gaussianity (PNG) has received in the literature,  we quantified the effects of PNG on the giant-arc abundance in \cite{2011MNRAS.415.1913D}.   Our work in \cite{2011MNRAS.415.1913D} focused on a widely used parameterization of PNG - the local form (e.g. \cite{2001PhRvD..63f3002K}) - in which the $\fNL$ parameter is constant.  Perhaps not surprisingly, we found that PNG of the local form can have only a modest effect within the most recent constraints from the Wilkinson Microwave Anisotrpy Probe (WMAP), which limit $-10< \fNL < 74$ at the $95 \%$ confidence level\cite{2011ApJS..192...18K}.  

However, non-standard inflationary scenarios can lead to a scale-dependent $\fNL$ which can have a larger impact on the scales relevant for galaxy cluster formation, while at the same time satisfying CMB and LSS constraints \cite{Lo-Verde:2008rt}.  Here we extend our calculations in \cite{2011MNRAS.415.1913D} to the parameterization proposed in \cite{Lo-Verde:2008rt}, where the primordial bispectrum has the equilateral shape\cite{2006JCAP...05..004C} and we make the replacement,

\begin{equation}
\fNL^{\mathrm{eq}} \rightarrow \fNL^{\mathrm{eq}} \left( \frac{k_1 + k_2 + k_3}{k_{\mathrm{CMB}}}  \right)^{-2 \kappa_{\mathrm{NG}}}.
\end{equation}   
From here on we refer to this parameterization as scale-dependent equilateral (SDE).  In what follows, we will vary the exponent $\kappa_{\mathrm{NG}}$ in order to explore various scale-dependent examples, but we will assume a fixed pivot wavenumber $k_{\mathrm{CMB}} = 0.086h~\Mpc^{-1}$, which approximately corresponds to the maximum multipole used in the WMAP year-seven analysis for constraining PNG \cite{2009MNRAS.394..133C}.   The current WMAP $95\%$ confidence limits for the equilateral shape are $-214 < \fNL^{\mathrm{eq}} < 266$\cite{2011ApJS..192...18K}.  We will assume that these values also apply for the SDE case at the pivot scale, even though actual SDE constraints would likely be even weaker \cite{Lo-Verde:2008rt}.

The probability for a background galaxy at redshift $z_s$ to produce giant arcs is given by the optical depth\cite{2003A&A...409..449B},
\begin{equation}
\tau(z_s) = \int_0^{z_s}{\mathrm{d}z \frac{\mathrm{d}V}{\mathrm{d}z}\int_{M_{\mathrm{min}}}^{\infty}{\mathrm{d}M~\frac{\mathrm{d}n}{\mathrm{d}M}~\sigma_{\mathrm{a}}(M,z)}},
\label{EQ:opticaldepth}
\end{equation} 
where $\sigma_{\mathrm{a}}$ is the giant-arc cross section\footnotemark, $\dd V/\dd z$ is the comoving volume element, $\dd n/\dd M$ is the halo mass function, and $M_{\mathrm{min}}$ is the minimum mass to produce giant arcs (see section 4.3 of \cite{2011MNRAS.415.1913D} for a discussion).  \footnotetext{Note that we have utilized the approximation of \cite{2003A&A...409..449B} for $\sigma_a$.  In this case, the cross section is in angular units.  Note that the angular diameter distance to $z_s$ does not appear in equation (\ref{EQ:opticaldepth}).} 

As in \cite{2011MNRAS.415.1913D}, we focus on two potential ways that PNG can influence the giant arc frequency.  First, PNG can affect the abundance of galaxy clusters, $\mathrm{d}n / \mathrm{d}M$, which would lead to a change in the number of supercritical lenses that are available in the appropriate redshift range.  To take into account the effects of PNG on the cluster abundance, we use the mass function of \cite{2011PhRvD..83d3526S}, which is based on the form originally derived by \cite{Lo-Verde:2008rt} (also see \cite{2010ApJ...717..526M}).  


\begin{figure}
\begin{center}
\resizebox{6.4cm}{!}{\includegraphics{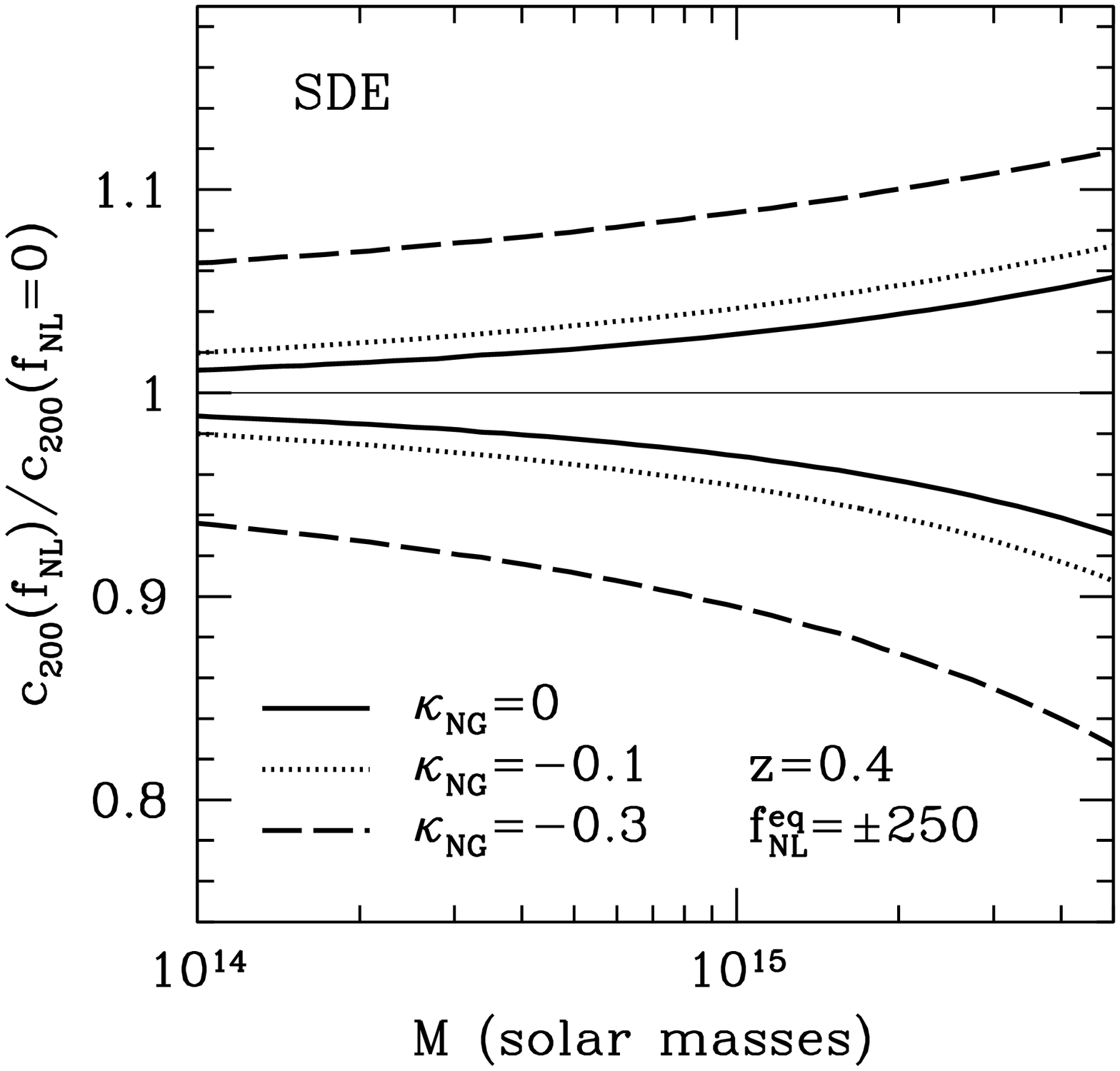}}
 \resizebox{6.4cm}{!}{\includegraphics{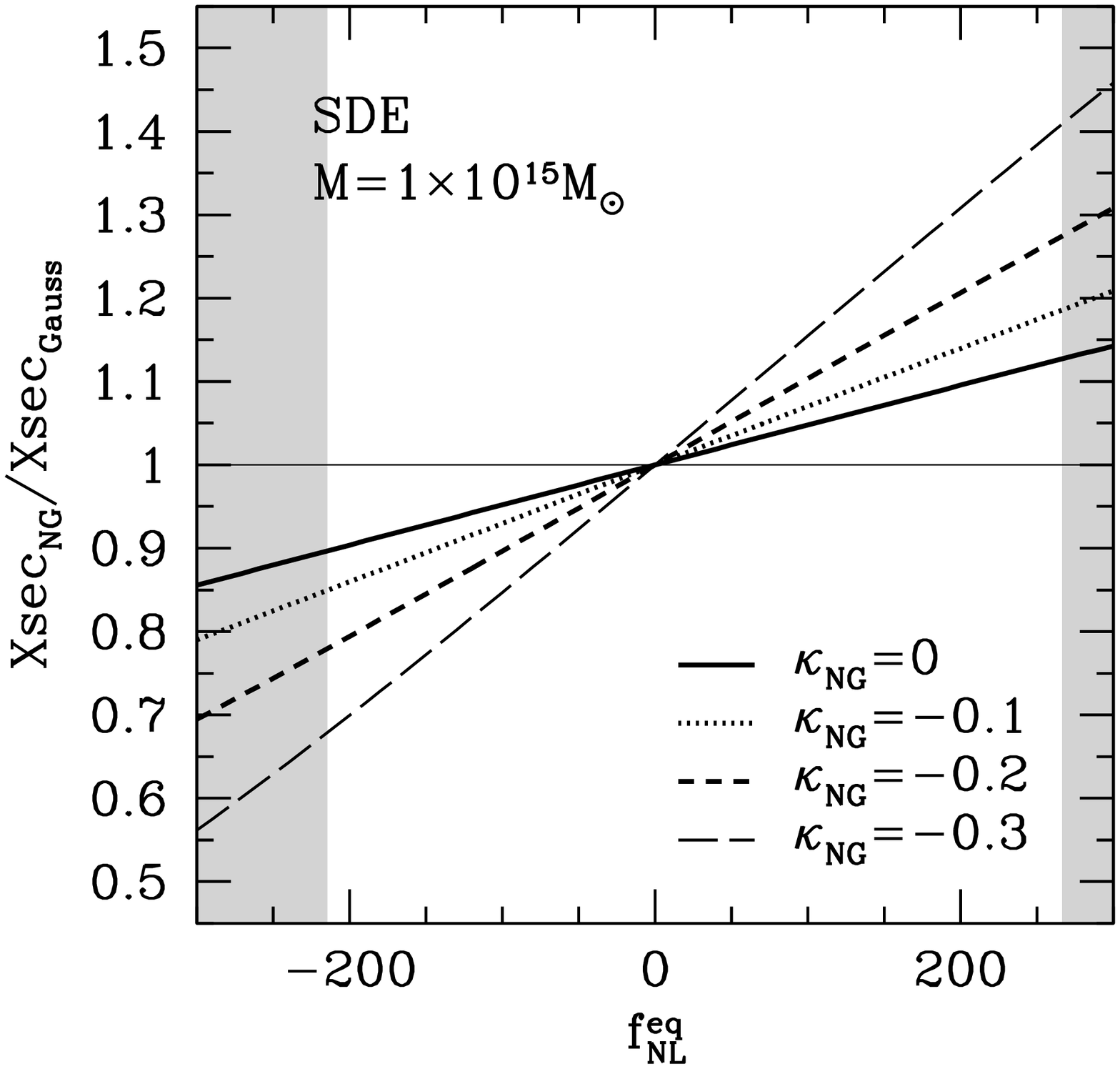}}
\end{center}
\caption{ Left panel:  the effect of SDE non-Gaussianity on mean halo concentrations.  Right panel:  resulting changes in giant-arc cross sections.  We assume $z_l=0.4$, $z_s=1.82$, and $\epsilon=0.2$. }
 \label{fig1}
 \end{figure}


Secondly, PNG is expected to influence the central densities of halos through its effect on the timing of structure formation (see \cite{2011MNRAS.415.1913D} and references therein).  Consider two model universes: one with Gaussian initial conditions and the other with non-Gaussian initial conditions (with $\fNL>0$ for concreteness).  In each universe, suppose we identified all halos with mass $M$ at redshift $z$, and compared the two sets of halos.  The set of halos in the universe where $\fNL>0$ would tend to have larger central densities compared to the Gaussian set.  We turn to a simple heuristic argument to understand this effect.    We may draw a rough correspondence between a halo with mass $M$ and a point in the linearly extrapolated density field where the density fluctuation reaches a threshold for collapse, $\delta_c$, when it is smoothed about that point on a scale corresponding to $M$.  As the smoothing scale is decreased, the conditional probability that the density fluctuation makes upward excursions is larger for $\fNL>0$, relative to the Gaussian case, due to the enhanced tail of the conditional probability density function.  Therefore, in the $\fNL>0$ case, one has to go to higher redshifts on average, relative to the Gaussian case, to reach the epoch at which the same fraction of the final mass was accumulated.  Since the central density of a halo reflects the cosmic mean density at the epoch of its formation (e.g. \cite{1997ApJ...490..493N}), we would therefore expect $\fNL>0$ to yield larger central densities on average, relative to the Gaussian case.   A similar argument leads to the opposite conclusion for $\fNL<0$.  In this case the formation epoch is delayed, and the central densities are lower.  

In \cite{2011MNRAS.415.1913D} we used techniques introduced by  \cite{2010ApJ...717..526M} to quantify the above effects and the resulting changes in mean halo concentrations.    Note that changes to central densities result in changes to the lensing cross sections, $\sigma_a$, and minimum mass threshold, $M_{\mathrm{min}}$, which appear in equation (\ref{EQ:opticaldepth}).  Here, we extend our calculations to the SDE case.  The left panel of Figure \ref{fig1} shows the ratio of non-Gaussian to Gaussian mean halo concentrations as a function of mass.  We use a fixed redshift $z=0.4$, corresponding to the redshift of typical cluster lenses.  The top and bottom set of curves correspond to $\fNLeq = 250$ and $\fNLeq = -250$ respectively.  In the right panel of figure \ref{fig1}, we show the resulting changes to the giant-arc cross sections. The shading corresponds to $\fNLeq$ values for the scale-independent equilateral shape excluded at the $95\%$ level by the WMAP year seven analysis.  We use a lens redshift of $z_l = 0.4$, $\epsilon = 0.2$, which describes the ellipticity of the lensing potential (see section 4.2 of \cite{2011MNRAS.415.1913D}), and a source redshift of $z_s = 1.82$, which is the median redshift observed in the Sloan Giant Arcs Survey \cite{2011ApJ...727L..26B}.

\begin{figure}
\begin{center}
\resizebox{6.4cm}{!}{\includegraphics{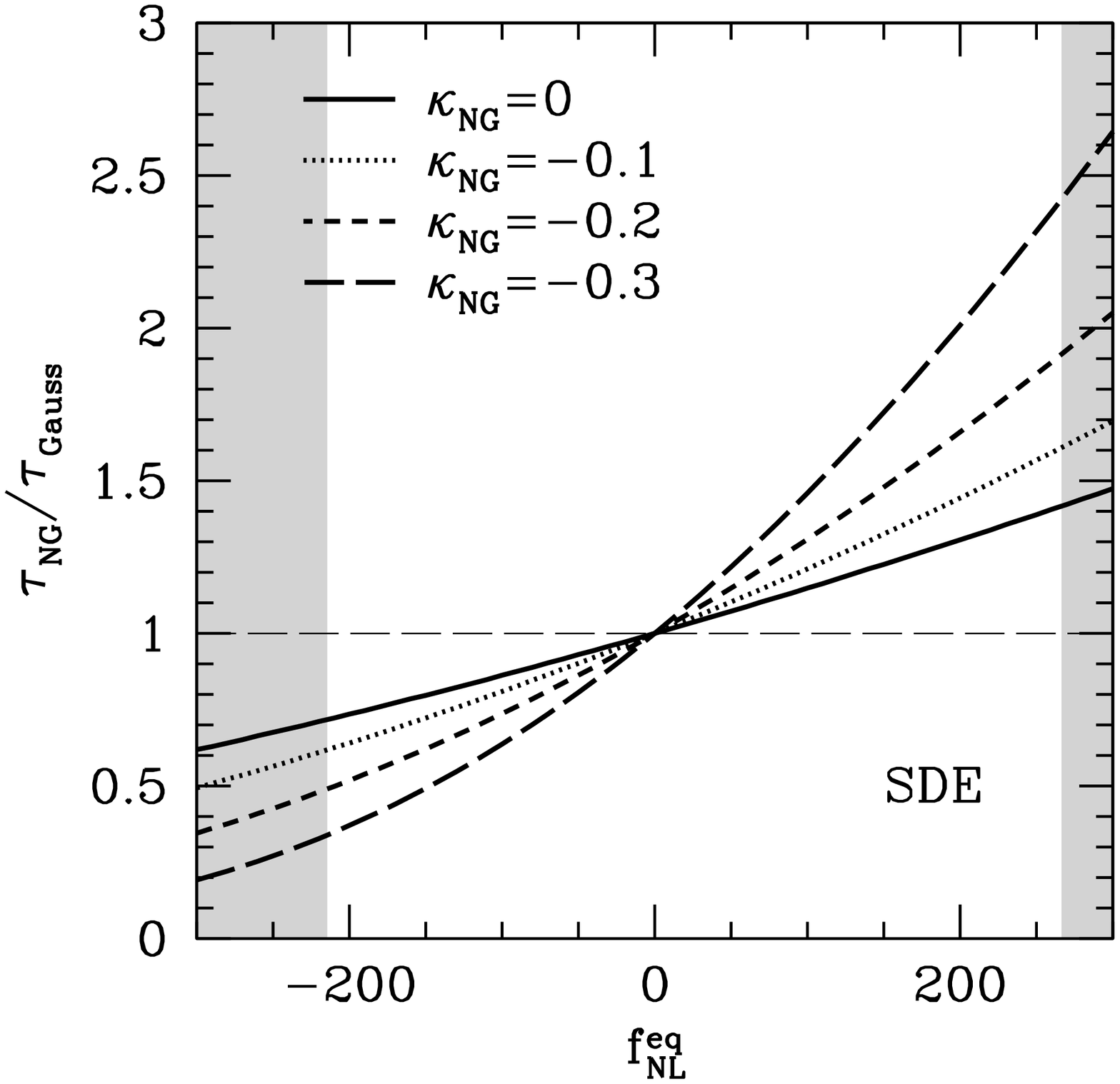}}
 \resizebox{6.4cm}{!}{\includegraphics{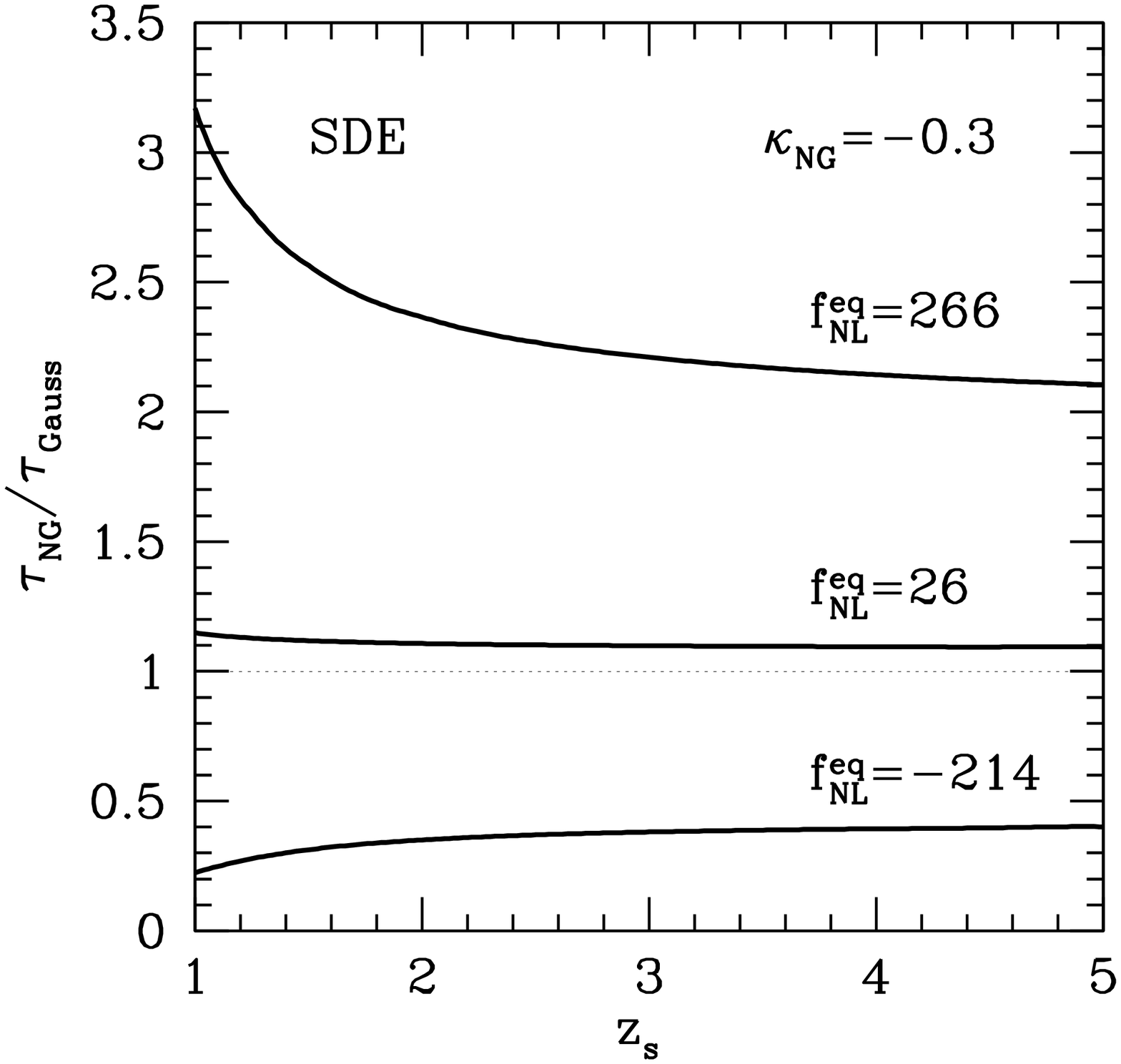}}
\end{center}
\caption{Relative changes in the giant-arc optical depth due to SDE non-Gaussianity.  We assume $\epsilon=0.2$ and $\theta_{\mathrm{min}}=10''$ (see \cite{2011MNRAS.415.1913D} for a discussion of these parameters).}
 \label{fig2}
 \end{figure}

The ratio of non-Gaussian to Gaussian giant-arc optical depths for $z_s=1.82$ is shown in the left panel of Figure \ref{fig2}.  The right panel of Figure \ref{fig2} shows the ratio as a function of $z_s$. We note that the deviations from the Gaussian case in $\tau$ are due to the combined effects of modified central densities and halo abundance.  For example, in the case with $\fNLeq > 0$, central densities are enhanced \emph{and} the abundance of large-mass halos is increased, which can boost the giant-arc optical depth substantially.  Note that PNG of the SDE type can, within the latest CMB constraints, yield up to a factor of a few difference in the optical depth.  Compare this to the maximum effect of a few tens of per cent found in \cite{2011MNRAS.415.1913D} for the local form.  


While our simple model allows us to quantify relative differences due to PNG, accurately predicting giant-arc abundances is well beyond the scope.  However, we can use our model to get ``back-of-the-envelope" estimates of what these changes imply in practice.   For this task, we use a fixed $\dd N_s / \dd z_s$ obtained from the observed galaxy redshift distribution in the Canada-France-Hawaii Telescope Legacy Survey \cite{2008A&A...479....9F}, and the all-sky extrapolation of roughly $1000$ arcs with length-to-width ratio $\ge10$ and R-band magnitudes $<21.5$ \cite{1994ApJ...422L...5L,1998A&A...330....1B,2004ApJ...609...50D}.  If we assume that the theoretical prediction for the Gaussian model is of order $\sim1000$ giant arcs, then the SDE non-Gaussian cases with $\kappa_{\mathrm{NG}}=-0.1$ and $\fNL^{\mathrm{eq}}=26(266)$ would predict 50(560) more giant arcs, whereas $\fNL^{\mathrm{eq}}=-214$ would lead to 360 less.  In the most extreme case considered here with $\kappa_{\mathrm{NG}}=-0.3$, $\fNL^{\mathrm{eq}}=26(266)$ would predict 100(1320) more giant arcs, while $\fNL^{\mathrm{eq}}=-214$ would yield 640 less 

In summary, within the latest CMB constraints, PNG of the local type can alter the giant-arc abundance by a maximum of a few tens of percent \cite{2011MNRAS.415.1913D}.  In this work, we have shown that non-standard scenarios with other bispectrum shapes and scale-dependent $\fNL$, such as the SDE model considered here, can modify the predicted giant-arc abundance by up to a factor of a few.





\bibliographystyle{siam}
\bibliography{PNGod}




\end{document}